\documentclass{pasa}%

\pdfoutput=1
\usepackage{graphicx}	
\usepackage{amssymb}	
\usepackage{epstopdf}
\usepackage{longtable}

\newcommand{\ms}{\mathrm{m\,s^{-1}}}

\newcommand{\msun}{M$_{\odot}$}

\newcommand{\rsun}{R$_{\odot}$}
\newcommand{\mstar}{$M_{\star}$}
\newcommand{\rstar}{$R_{\star}$}
\newcommand{\lstar}{$L_{\star}$}
\newcommand{\teff}{$T_{{\rm eff}}$}
\newcommand{\logg}{\ensuremath{\log{g}}}
\newcommand{\feh}{\ensuremath{{\rm [Fe/H]}}}
\newcommand{\vsini}{\ensuremath{v \sin{i}}}

\title[Detailed analysis of HD\,76920\,b]{HD\,76920\,b pinned down: a detailed analysis of the most eccentric planetary system around an evolved star}

\author[C. Bergmann et al.]{
C. Bergmann$^{1,2}$\thanks{Visiting Astronomer, University of Canterbury Mt John Observatory}\,\,\,\thanks{E-mail: cbergmann001@gmail.com}, 
M. I. Jones$^{3,4}$,
J. Zhao$^{1,5}$,
A. J. Mustill$^{6}$,
R. Brahm$^{7,8}$,
P. Torres$^{9}$,
R. A. Wittenmyer$^{10}$,
F. Gunn$^{11}$,
K. R. Pollard$^{11}$,
A. Zapata$^{9}$,
L. Vanzi$^{9}$,
S. Wang$^{12}$
\affil{$^1$Exoplanetary Science at UNSW, School of Physics, UNSW Sydney, NSW 2052, Australia}%
\affil{$^2$Deutsches Zentrum f\"{u}r Luft- und Raumfahrt, M\"{u}nchener Str. 20, 82234 We{\ss}ling, Germany}%
\affil{$^3$European Southern Observatory, Alonso de C\'ordova 3107, Casilla 19001, Santiago, Chile}%
\affil{$^4$Instituto de Astronom\'ia, Universidad Cat\'olica del Norte, Angamos 0610, 1270709, Antofagasta, Chile}
\affil{$^5$Penn State University, Department of Astronomy and Astrophysics, University Park, PA 16802, USA }
\affil{$^6$Lund Observatory, Department of Astronomy \& Theoretical Physics, Lund University, Box 43, 221 00 Lund, Sweden}%
\affil{$^7$Millennium Institute for Astrophysics, Chile}
\affil{$^8$Facultad de Ingenier\'ia y Ciencias, Universidad Adolfo Ib\'a\~nez, Av. Diagonal las Torres 2640, Pe\~nalol\'en, Santiago, Chile}
\affil{$^9$Department of Electrical Engineering and Center of Astro Engineering, Pontificia Universidad Cat\'olica de Chile, Av. Vicu\~na Mackenna 4860, Santiago, Chile}%
\affil{$^{10}$University of Southern Queensland, Centre for Astrophysics, Toowoomba, Queensland 4350, Australia}%
\affil{$^{11}$School of Physical and Chemical Sciences, Te Kura Mat\={u}, University of Canterbury, Christchurch 8020, New Zealand}%
\affil{$^{12}$Department of Astronomy, Indiana University, Bloomington, IN 47405, USA}%
}%

\jid{PASA}
\doi{10.1017/pas.\the\year.xxx}
\jyear{\the\year}

\usepackage{aas_macros}
\usepackage{hyperref} 
\hypersetup{colorlinks,citecolor=blue,linkcolor=blue,urlcolor=blue}

\hypersetup{draft}

\begin{document}

\begin{frontmatter}
\maketitle

\begin{abstract}
We present 63 new multi-site radial velocity measurements of the K1III giant HD\,76920, which was recently reported to host the most eccentric planet known to orbit an evolved star. We focussed our observational efforts on the time around the predicted periastron passage and achieved near-continuous phase coverage of the corresponding radial velocity peak. By combining our radial velocity measurements from four different instruments with previously published ones, we confirm the highly eccentric nature of the system, and find an even higher eccentricity of $e=0.8782 \pm 0.0025$, an orbital period of $415.891^{+0.043}_{-0.039}\,\mathrm{d}$, and a minimum mass of $3.13^{+0.41}_{-0.43}\,\mathrm{M_J}$ for the planet. The uncertainties in the orbital elements are greatly reduced, especially for the period and eccentricity. We also performed a detailed spectroscopic analysis to derive atmospheric stellar parameters, and thus the fundamental stellar parameters ($M_*, R_*, L_*$), taking into account the parallax from Gaia DR2, and independently determined the stellar mass and radius using asteroseismology. Intriguingly, at periastron the planet comes to within 2.4 stellar radii of its host star's surface. However, we find that the planet is not currently experiencing any significant orbital decay and will not be engulfed by the stellar envelope for at least another $50-80$\,Myr. Finally, while we calculate a relatively high transit probability of $16\%$, we did not detect a transit in the TESS photometry.
\end{abstract}

\begin{keywords}
planetary systems -- stars: individual: HD\,76920 -- techniques: radial velocities
\end{keywords}
\end{frontmatter}

\section{INTRODUCTION}
\label{sec:intro}

Since the first detection of a planet around a main-sequence star more than two decades ago \citep{51peg}, thousands of planetary systems have been found with astonishing diversity. The transit method has produced the most planet discoveries to date, mostly due to the overwhelming success of NASA's \textit{Kepler} mission \citep{kepler}. Yet, a significant fraction of all planets has been found using the radial velocity (RV) technique, including many longer-period planets, for which ground-based Doppler searches have a natural advantage over short-lived space missions. The RV method also remains a valuable tool for the detection and characterization of planets, as it yields the minimum mass of a planet and therefore nicely complements the transit method, which yields the size of a planet. Space-based transit searches such as \textit{Kepler} \citep{kepler} or TESS \citep{tess} also rely on ground-based RV follow-up observations to confirm their planet candidates, to determine their masses and hence densities, and to search for additional planets in these systems. 

Slowly rotating solar-type and late-type stars are ideal targets for Doppler planet searches, as their spectra exhibit numerous sharp absorption lines. However, main-sequence stars more massive than about $1.5\,\mathrm{M_\odot}$ are generally not suitable for precise RV  measurements, due to their high temperatures and fast rotation rates \citep[e.g.][]{galland2005}. Consequently, the occurrence rate and distribution of planets as a function of stellar mass and metallicity for intermediate-mass stars are not as well-established compared to planets around lower-mass stars \citep[e.g.][]{johnson2010,planfreq_lsw,jones2016,ppps6}. In order to learn more about planetary systems around these intermediate-mass stars, several planet search programmes have been specifically targeting evolved stars of the same mass (often dubbed 'retired A-stars'), as they do not suffer from these effects \citep[e.g.][]{iota_dra,sato2003,retired_astars,ps_torun,ppps,jones_giants}. 

More than 100 planets have now been discovered orbiting giant stars, and except for the very few such systems discovered by \textit{Kepler} \cite[e.g.][]{kepler432}, all of these have been discovered via the RV technique. Notably the first planet found to orbit a giant star, $\iota$\,Dra\,b, also happens to be on a very eccentric orbit with $e = 0.71$ \citep{iota_dra,iotadra_transitprob}. More recently, planets on even more eccentric orbits were found around the K3III giant BD+48\,740 (HIP\,12684) with $e = 0.76$ \citep{adamow2012,adamow2018}, and around the K1III giant HD\,76920 with $e = 0.86$ \citep{w17}. The latter currently claims the title of being the most eccentric planet orbiting an evolved star, and is the subject of this work.

Orbits with certain combinations of high eccentricities and longitudes of periastron produce RV curves that are essentially flat for the majority of the orbital period and only exhibit a narrow peak near periastron passage. However, because nearly all the information needed to determine the orbital parameters, including the RV semi-amplitude and thus the minimum mass of the planet, is contained in this short phase of the orbit, it is rather difficult from an observational point of view to obtain a good orbital solution for such systems. In their discovery paper, \citet{w17} pointed out that the periastron passage of HD\,76920\,b still needed better observational sampling in order to sufficiently constrain the orbital elements and minimum mass of the planet. We therefore planned a multi-site observing run and successfully filled in the gap in the phase coverage near periastron passage with RV measurements. 

In this paper we present the results of our observational campaign. We describe our multi-site observations and RV measurements in Sections~\ref{sec:obs} and \ref{sec:rvs}, respectively. We then present newly computed stellar parameters in Section~\ref{sec:stellar_parms}. In Section~\ref{sec:solution} we describe the orbit fitting process and present our improved orbital solution. We also present an upper mass limit for HD\,76920\,b in Section~\ref{sec:mass_limit}, followed by a discussion of our findings in Section~\ref{sec:discussion}.

\section{OBSERVATIONS}
\label{sec:obs}
Between January and March 2018, we acquired 63 observations of HD\,76920 using four different instruments. Firstly, 39 were taken at the University of Canterbury Mt John Observatory (UCMJO) in Lake Tekapo, New Zealand, using the 1-m McLellan telescope in conjunction with the HERCULES spectrograph \citep{hercules}. Of these, 19 were taken with a 100-$\mu$m core-diameter fibre ('fibre 1'), corresponding to a resolving power of $R \sim 41\,000$, and another 20 were taken with a different 100-$\mu$m core-diameter fibre with a 50-$\mu$m micro-slit attached to its end ('fibre 3'), corresponding to a resolving power of $R \sim 70\,000$. These observations are referred to below as MJ1 and MJ3, respectively. 

A further 8 observations were obtained with the CHIRON spectrograph \citep{chiron} attached to the 1.5-m telescope at Cerro Tololo Inter-American Observatory (CTIO) in Chile. These new spectra were obtained using the 'slit mode', which delivers a resolving power of $R \sim 95\,000$ and a total system efficiency of $\sim$ 2\%. Both HERCULES and CHIRON employ the iodine-cell technique \citep[e.g.][]{3ms_precision, austral}, i.e. during the observations a gas absorption cell containing molecular iodine (kept at a constant temperature of $50.0 \pm 0.1^\circ\,\mathrm{C}$) is placed in the light path for wavelength calibration and for modelling of the instrumental profile. 

Finally, we also obtained 15 spectra using FIDEOS \citep{fideos} and one new spectrum using FEROS \citep{feros}. These two high-resolution ($R \sim 43\,000$ and $R \sim 48\,000$, respectively) spectrographs are located at La Silla Observatory in Chile, and are attached to the ESO 1-m and 2.2-m telescopes, respectively. In addition, they are both fed by multiple optical fibres, allowing a simultaneous ThAr wavelength calibration during the science exposure for precision radial velocities.

\section{RADIAL VELOCITIES}
\label{sec:rvs}
Raw-reduction of the HERCULES observations was performed with the latest version (v5.2.9) of the HERCULES Reduction Software Package \citep[\texttt{HRSP};][]{hrsp} and the pipeline described in \citet{cmb_thesis}. From the reduced spectra we derived radial velocities using our version of the \texttt{AUSTRAL} Doppler code described by \citet{austral}. We used a high-S/N, iodine-free spectrum of the K1III star $\nu$\,Octantis as a template \citep{ramm2015}. While this setup has proven to deliver long-term RV precisions of about $4.5\,\ms$ (with short-term precision of $\lesssim 3\,\ms$) for bright solar-type stars \citep{cmb_thesis, cmb_ijasb}, the fainter magnitude of HD\,76920 $(V=7.8)$ combined with poor seeing conditions for a majority of the nights resulted in single-shot uncertainties of typically $14.5\,\ms$ for the higher-resolution fibre, and $16.5\,\ms$ for the lower-resolution fibre, which was preferred if the seeing was worse than about 3 arcsec.

The CHIRON spectra were reduced with the observatory customized pipeline, which provides order by order extracted and wavelength calibrated spectra for CHIRON users. The RVs were computed following the method described in \citet{jones2017}. We note that the new CHIRON velocities have larger uncertainties compared to those in \citet{w17}, as the new spectra were obtained in 'slit mode', rather than in 'fibre-slicer mode' ($R \sim 80\,000$). As a consequence, although the new data were taken at a slightly higher resolving power, the efficiency is drastically reduced (by a factor of $\sim$ 3), when compared to 'fibre-slicer mode', directly translating into lower S/N data, and thus leading to larger RV uncertainties. We have also recomputed the 'fibre-slicer mode' CHIRON RVs published in \citet{w17}, including the new data, which resulted in small but non-negligible changes. 

Finally, both FIDEOS and FEROS data were processed with the \texttt{CERES} code \citep{ceres_code}, including a re-reduction of the 8 FEROS observations published in \citet{w17}. \texttt{CERES} performs a standard \'{e}chelle spectra reduction including bias subtraction, order tracing, optimal extraction, and wavelength calibration. The RVs for the two instruments were obtained from the cross-correlation function \citep[CCF;][]{ccf_rvs}. In the case of FIDEOS, the template used for the CCF corresponds to a numerical binary mask as explained in \citet{ceres_code}, while in the case of FEROS data we use a high S/N template, which is built by stacking all of the individual observed spectra \citep{jones2017}. The final velocities are obtained after correcting for the night drift (from the simultaneous calibration fibre) and barycentric velocity. Note that the newly computed FEROS RVs are superior to those presented in \citet{w17}, where a much lower S/N template was used. All radial velocities used in this work and their corresponding uncertainties are summarized in Table \ref{tab:rvs}.

 \section{Stellar parameters}
 \label{sec:stellar_parms}
 \subsection{Spectroscopy}
 \label{sec:spectroscopy}
 We computed the atmospheric parameters of HD\,76920 using the \texttt{ZASPE} code \citep{zaspe_code}. For this purpose, we first combined all of the individual FEROS spectra. The co-added master spectrum is then compared to the ATLAS9 grid of stellar models \citep{atlas}, in carefully selected regions that are more sensitive to changes in the atmospheric parameters. This procedure is performed iteratively until we obtain the effective temperature (\teff), the surface gravity (\logg), the stellar metallicity (\feh), and the projected rotational velocity (\vsini). Using these derived parameters, we then computed the corresponding spectral energy distribution (SED). For this we used the \texttt{BT-Settl-CIFIST} models \citep{baraffe2015}. From the SED we computed synthetic magnitudes and we compared them to the observed ones, which are listed in Table \ref{tab:star_par}. During this process the stellar radius (\rstar) and the visual extinction ($A_V$) are derived, and thus the stellar luminosity (\lstar).  
 
\begin{figure}
\includegraphics[width=.50\textwidth]{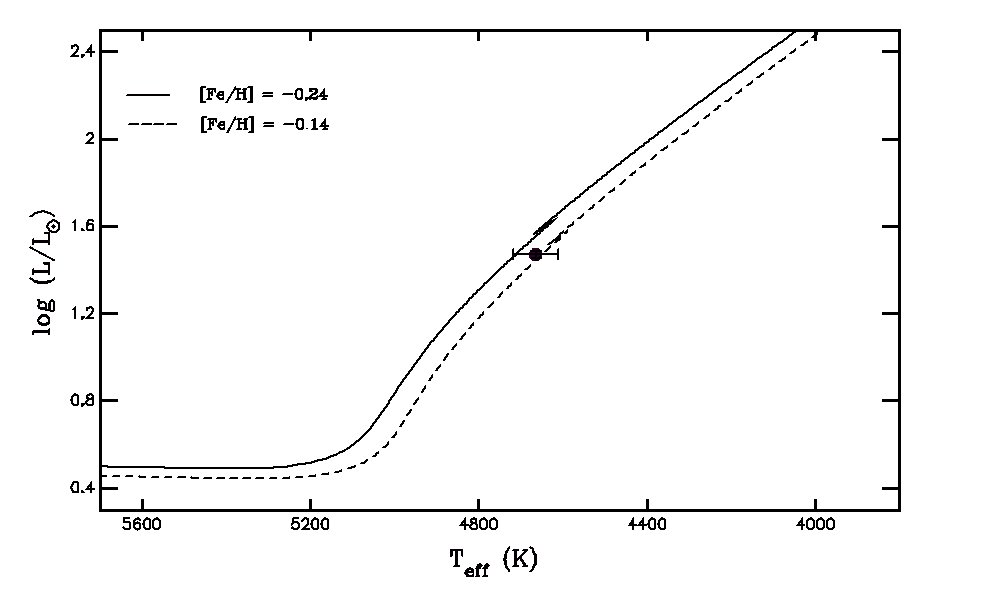}
\caption{HR diagram showing the position of HD\,76920. The solid and dashed lines correspond to \texttt{PARSEC} models with \mstar$\, = 1.0\,$\msun, for \feh$\,= -0.24$ and $-0.14\,\mathrm{dex}$, respectively.}
\label{fig:HRD}
\end{figure}

 Finally, to obtain the stellar mass and evolutionary status of HD\,76920, we compared the derived atmospheric parameters to the \texttt{PARSEC} evolutionary models \citep{parsec}. We found that HD\,76920 has a mass of \mstar$\,= 1.0 \pm 0.2$\,\msun, and that it is ascending the red giant branch (RGB) phase. Figure \ref{fig:HRD} shows the position of HD\,76920 in the HR diagram. For comparison, two different \texttt{PARSEC} isomass evolutionary tracks are overplotted. As can be seen, HD\,76920 is located midway on its RGB ascent, and is reaching the luminosity bump region. We note that no horizontal branch track (i.e. Helium burning core) crosses its position in the HR diagram. All derived atmospheric and physical parameters are listed in Table \ref{tab:star_par}.
 
 \subsection{Asteroseismology}
 \label{sec:asteroseismology}
\begin{figure}
\includegraphics[width=.5\textwidth]{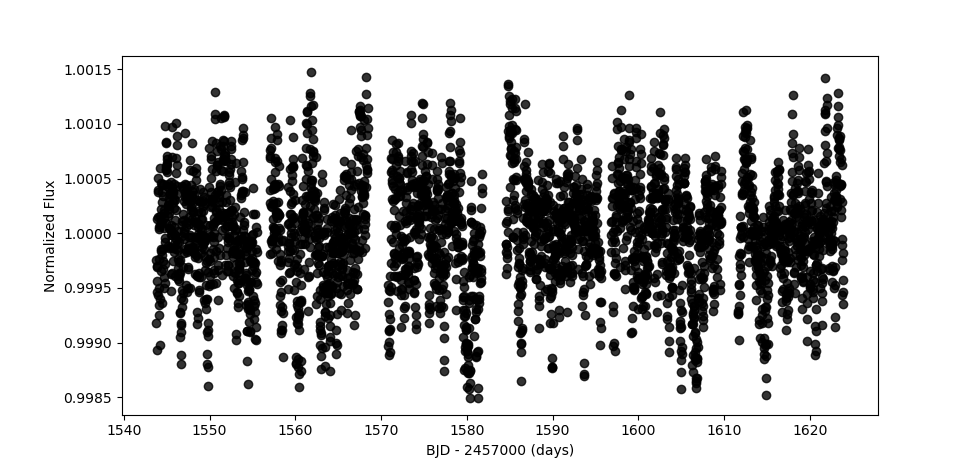}
\caption{Normalized and detrended TESS photometry of HD\,76920.}
\label{fig:tess_phot}
\end{figure}

 As an independent method, we used asteroseismology to derive the stellar mass from the TESS \citep{tess} photometric data. The TESS mission observed HD\,76920 in the long cadence mode (30 min) in sectors 9, 10, and 11, adding up to a total of 3492 individual photometric measurements. To obtain the light curve we used the python tool \texttt{tesseract} (Rojas et al. in prep) using the \texttt{autoap} aperture. We removed the most deviant outliers using the \texttt{clean.py} tool and normalized each sector data independently by its median value. We also detrended the light curve using a linear fit, achieving a dispersion of 494.6 ppm. Figure \ref{fig:tess_phot} shows the resulting normalized TESS photometry of HD\,76920. Then we ran a generalized Lomb-Scargle (GLS, \cite{gls_paper}) routine to obtain the power spectral density (PSD) and search for asteroseismic power excess in order to measure $\nu_{\rm max}$ and $\Delta\nu$. We also corrected the background using a very wide gaussian profile kernel. After this correction we followed the method presented in \citet{jones2018}, which consists of convolving a gaussian profile with a $\sigma= 12\mu {\rm Hz}$ kernel around the power excess in order to find the peak that corresponds to $\nu_{\rm max}$. To obtain $\Delta\nu$ we convolved a gaussian profile with a $\sigma= 1\mu {\rm Hz}$ kernel and we ran an autocorrelation routine. Using this procedure, we obtained $\nu_{\rm max} = 54.01 \pm 2.75\,\mu {\rm Hz}$  and $\Delta\nu = 5.64 \pm 0.24\,\mu {\rm Hz}$. From these values, and following the scaling relations presented in \citet{scalingrelations}, we obtained a mass of \mstar$\,= 1.29 \pm 0.17\,$\msun\, and a radius \rstar$\,= 9.07 \pm 0.63$\,\rsun . The corresponding 1-$\sigma$ error bars were obtained from a bootstrap analysis. The asteroseismic mass and radius are in reasonably good agreement with the spectroscopic values.

\begin{table}
	\centering
	\caption{Stellar parameters of HD\,76920}
	\label{tab:star_par}
	\renewcommand{\arraystretch}{1.2} 
		\begin{tabular}{lcc}
		\hline
Parameter               &     Value            & Method/Source \\
\hline
B\,\,[mag]             &  8.83 $\pm$ 0.02      &  Tycho-2 \\
V\,\,[mag]             &  7.82 $\pm$ 0.01      &  Tycho-2 \\
G\,\,[mag]             &  7.5144 $\pm$ 0.0003  &  Gaia   \\
J\,\,[mag]             &  5.95 $\pm$ 0.02      &  2MASS  \\
H\,\,[mag]             &  5.41 $\pm$ 0.04      &  2MASS  \\ 
K\,\,[mag]             &  5.25 $\pm$ 0.03      &  2MASS  \\ 
W1\,\,[mag]            &  5.187 $\pm$ 0.197    &  WISE   \\
W2\,\,[mag]            &  5.097 $\pm$ 0.062    &  WISE    \\
W3\,\,[mag]            &  5.201 $\pm$ 0.014    &  WISE    \\

\vspace{-0.5cm} \\
\vspace{-0.3cm}                                                      \\		
\teff\,\,[K]            &  4664 $\pm$ 53            & \texttt{ZASPE}           \\
\logg\,\,[cm\,s$^{-1}$] &  2.71 $\pm$ 0.04          & \texttt{ZASPE} + Gaia    \\
\feh\,\,[dex]           & -0.19 $\pm$ 0.06          & \texttt{ZASPE}            \\
\vsini\,\,[km\,s$^{-1}$]   &   2.5 $\pm$ 0.3   & \texttt{ZASPE}            \\ \vspace{-0.5cm} \\
\rstar\,\,[\rsun]      &  $8.33 \pm 0.56$  &  \texttt{ZASPE} + Gaia   \\
                       &  $9.07 \pm 0.63$    &  Asteroseis. + TESS \\
\lstar\,\,[L$_\odot$]  &  $29.5\,^{+1.3}_{-1.0}$     & \texttt{ZASPE} + Gaia \\
\mstar\,\,[\msun]      &  $1.0 \pm 0.2$            & \texttt{PARSEC}  \\
                       &  $1.29 \pm 0.17 $          &  Asteroseis. + TESS \\ \\ \hline
		\end{tabular}
\end{table}

\section{Orbital Solution}
\label{sec:solution}
\citet{w17} published 37 RVs of HD\,76920 obtained with three different spectrographs. Of these, 17 were taken with UCLES \citep{ucles} installed at the 3.9-m Anglo-Australian Telescope (AAT), 12 were taken with CHIRON at the 1.5-m telescope at CTIO, and 8 were taken with FEROS installed at the 2.2-m telescope at La Silla.

\begin{figure}
	\includegraphics[width=\columnwidth]{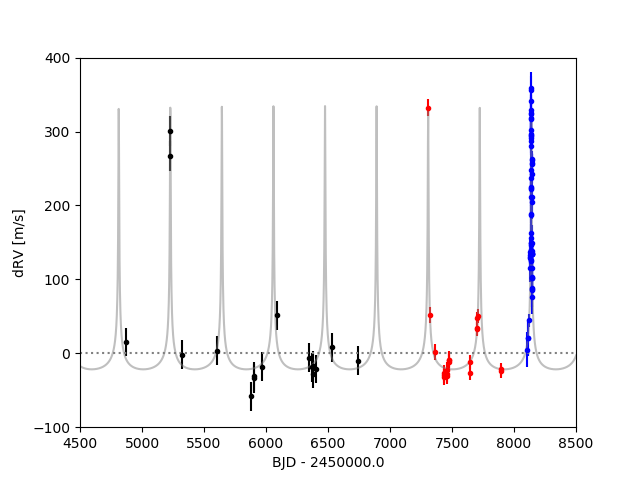}
	\caption{All available RV data together with our best-fit Keplerian orbital solution for HD\,76920\,b. The black points are the AAT data from \citet{w17}, the red points are the re-reduced CHIRON and FEROS data, and the blue points are the CHIRON, FEROS, FIDEOS, and HERCULES data taken for this work. Error bars represent the total uncertainty given by Eq.\,\ref{eq:errors}. The RMS about this fit is $14.1\,\ms$.}
	\label{fig:all_rvs}
\end{figure}

\begin{figure}
	\includegraphics[width=\columnwidth]{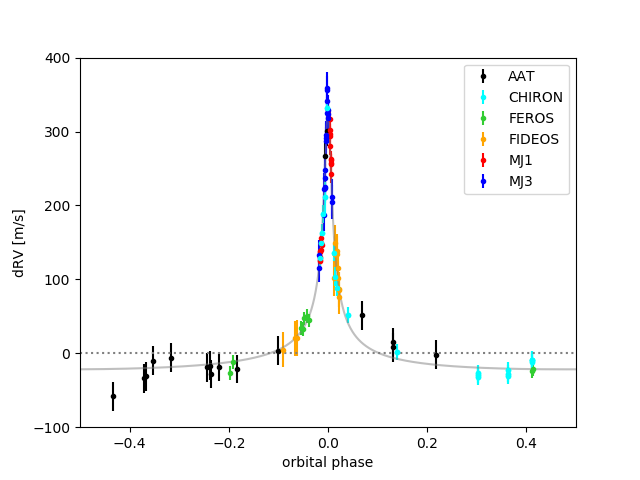}
	\caption{Same as Fig.\,\ref{fig:all_rvs}, but phase-folded on the orbital period. Data from different instruments/setups are shown in different colours: AAT\,\textemdash\,black, CHIRON\,\textemdash\,cyan, FEROS\,\textemdash\,green, FIDEOS\,\textemdash\,orange, MJ1\,\textemdash\,red, MJ3\,\textemdash\,blue.}
	\label{fig:all_rvs_phased}
\end{figure}

\begin{figure}
	\includegraphics[width=\columnwidth]{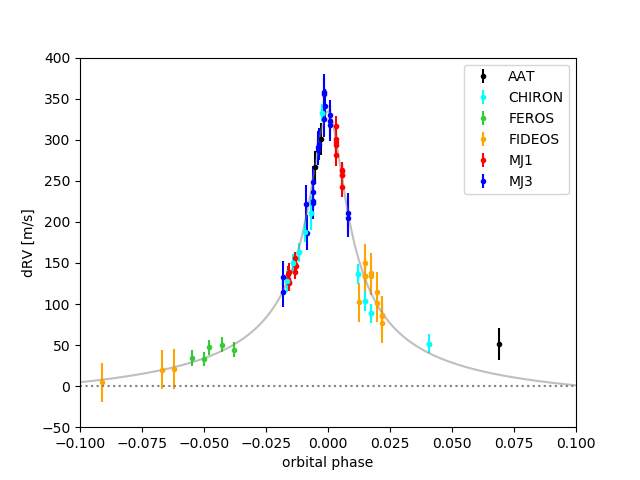}
	\caption{Close-up view of the RV peak near periastron passage. Colour-coding is the same as in Fig.\,\ref{fig:all_rvs_phased}.}
	\label{fig:all_rvs_phased_zoom}
\end{figure}

We combined our 63 new RV measurements with the 12 re-reduced CHIRON RVs, the 8 newly reduced FEROS RVs, and the remaining 17 UCLES RVs from \citet{w17}, and fitted a single Keplerian model to the combined data set consisting of a total of 100 RVs. We used the \texttt{emcee} 2.2.1 package \citep{emcee}, a pure-python implementation of the Affine-Invariant Markov chain Monte Carlo (MCMC) Ensemble sampler \citep{GW2010}, to obtain the best-fit parameters. We used logarithmic priors for $P$, $T_0$, and $K$, i.e. we fit in terms of $\log{P}$, $\log{T_0}$, and $\log{K}$, and used uniform priors for all other parameters\footnote{For $T_0$, which is technically allowed to assume negative values, this choice of parameterization introduces an unintentional, but weak informative prior on the parameter. However, as the occurrence of $T_0$ is periodic in uniform space, if fitted with a uniform prior, the posterior distribution may look bi-modal if the initial value is not optimal. Having a logarithmic prior makes the walkers converge to the optimal value quickly, as a linear change in $\log(T_0)$ results in an exponential change in $T_0$. Because $T_0$ is periodic, one positive solution is enough to derive all other solutions.}. We deployed 32 walkers and ran 3\,000 steps for the first burn-in phase until the walkers had explored the parameter space sufficiently. At the end of the first burn-in phase, walkers are re-sampled around the most probable position to reject bad samplings. We then continued with a second burn-in phase for another 3\,000 steps. All parameters have clearly converged after the second burn-in phase. Finally, we collected the samples from the last 10\,000 steps to calculate the maximum-likelihood set of parameters and estimate the uncertainties. The random zero-point offsets between the different instrumental setups were included as additional free parameters in the fitting process. For consistency with the work of \citet{w17}, we also added $7\,\ms$ of stellar jitter in quadrature to the error bars, as is appropriate for this type of star \citep{scalingrelations}, and as explained in \citet{wittenmyer2016}. Finally, following the method described by \citep{baluev}, we also included an ``instrumental jitter'' term in the fitting, which acts to ensure that the uncertainties from the different instrumental setups do not introduce a spurious weighting of the RV points. This can happen if the internal error bars are overestimated or underestimated for some instruments. In our case the internal error bars of the AAT and FIDEOS data seem to be substantially underestimated as indicated by the large positive values of the square of their instrumental jitter terms in Table\,\ref{tab:orbel}, whereas the MJ1 error bars seem to be somewhat overestimated as indicated by the negative value of the square of their instrumental jitter term in Table\,\ref{tab:orbel}. The total uncertainty for each RV measurement is thus given by 
\begin{equation}
    \sigma = \sqrt{\sigma_{\mathrm{int}}^2 + \sigma_{\mathrm{st.jitt.}}^2 + \sigma_{\mathrm{inst.jitt.}}^2} \quad ,
    \label{eq:errors}
\end{equation}
where $\sigma_{\mathrm{int}}$ is the internal error as reported in Table\,\ref{tab:rvs}, $\sigma_{\mathrm{st.jitt.}}$ is the stellar jitter, and $\sigma_{\mathrm{inst.jitt.}}$ is the instrumental jitter. The RMS around the combined fit is $14.1\,\ms$, and the weighted RMS is $11.7\,\ms$. Figure~\ref{fig:all_rvs} shows all data points together with our best-fit orbital solution, Fig.~\ref{fig:all_rvs_phased} shows a phase-folded version of that plot, and Fig.~\ref{fig:all_rvs_phased_zoom} shows a close-up view of the RV peak near periastron passage (again phase-folded with the orbital period). A corner plot of the posterior probability distributions of the parameters, generated from the last 10\,000 steps and demonstrating that all parameters are well constrained, is shown in Fig.~\ref{fig:corner_plot} in the appendix.

In order to confirm our MCMC results, we also fitted a single Keplerian orbit using the IDL package \texttt{RVLIN} \citep{rvlin}. Here we estimated the corresponding uncertainties in the orbital parameters via the bootstrapping algorithm from the \texttt{BOOTTRAN} package \citep{boottran} using 100\,000 steps. Because the extra instrumental jitter term cannot be set as a free parameter, we used the modified error bars given by Eq.\,\ref{eq:errors} as input to the fitting. While the uncertainty estimates derived with the bootstrapping method are somewhat larger, the two sets of best-fit parameters are in good agreement. Both best-fit single-Keplerian orbital solutions and the corresponding parameter uncertainty estimates are summarized in Table~\ref{tab:orbel}. 

For the calculation of the semi-major axis and minimum mass of the planet we give the values using both the spectroscopic stellar mass of $M_* = 1.0 \pm 0.2$\,\msun\, (see Sect.\,\ref{sec:spectroscopy}), as well as the asteroseismic mass of $1.29 \pm 0.17$\,\msun\, (see Sect.\,\ref{sec:asteroseismology}). For the rest of this work we will adopt the spectroscopically derived stellar mass and radius unless otherwise mentioned. Note that the uncertainty in the semi-major axis is completely dominated by the uncertainty in stellar mass, which to a lesser extent also affects the uncertainty in the minimum mass of the planet. Note that \citet{w17} used a mass of $1.17 \pm 0.20\,$\msun, which has a comparable relative error. This leads us to believe that their uncertainty estimate for $m_{\mathrm{P}}\,\sin i$ is underestimated and may not include the uncertainty in stellar mass.

\begin{figure}
	\includegraphics[width=\columnwidth]{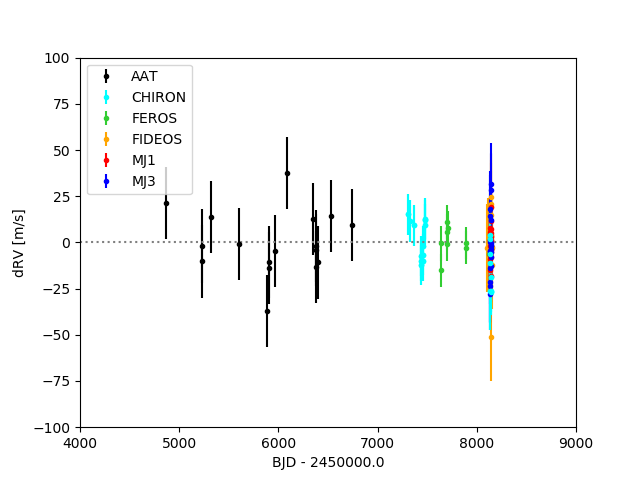}
	\caption{Residuals from the best-fit orbital solution. Error bars represent the total uncertainty given by Eq.\,\ref{eq:errors}. Colour-coding is the same as in Fig.\,\ref{fig:all_rvs_phased} and Fig.\,\ref{fig:all_rvs_phased_zoom}.}
	\label{fig:all_res}
\end{figure}

\begin{figure}
	\includegraphics[width=\columnwidth]{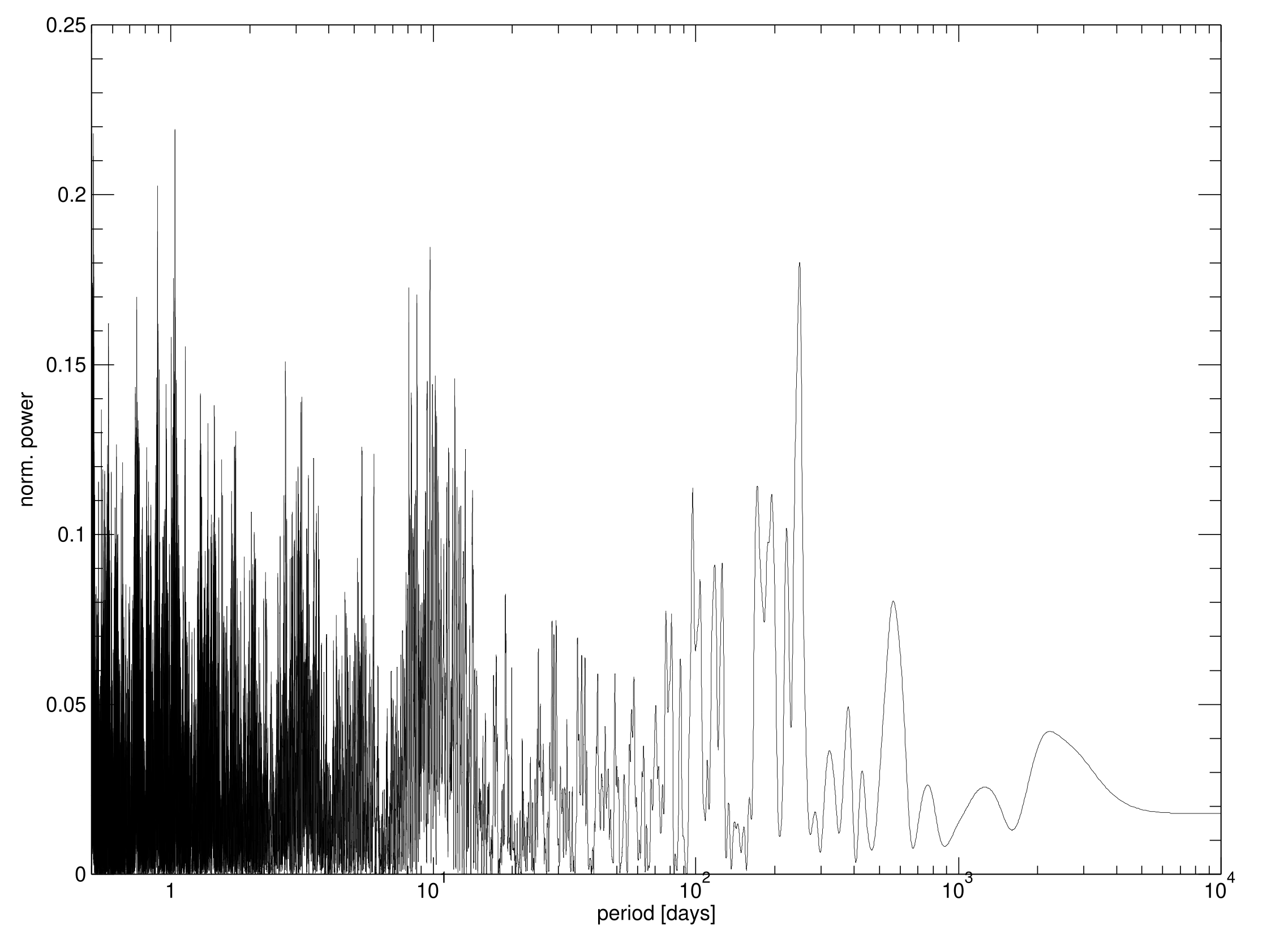}
	\caption{GLS periodogram of the residuals from the best-fit orbital solution shown in Fig.\,\ref{fig:all_res}.}
	\label{fig:gls}
\end{figure}

We also searched for periodic signals in the residuals (Fig.~\ref{fig:all_res}), but did not find any significant power, as can be seen in the GLS periodogram shown in Fig.~\ref{fig:gls}.

\begin{table*}
	\centering
	\caption{Best-fit orbital solution and derived quantities for HD\,76920\,b.\\
	$^\dagger$ A negative value for the square of the instrumental jitter indicates that the formal internal errors are overestimated and that the fitted instrumental jitter needs to be subtracted in quadrature in Eq.\,\ref{eq:errors}.}
	\label{tab:orbel}
	\renewcommand{\arraystretch}{1.3} 
		\begin{tabular}{lccc}
			\hline
			Element & \texttt{emcee} & \texttt{RVLIN} & W17\\\hline
			$P\,\mathrm{[d]}$ & $415.891^{+0.043}_{-0.039}$ & $415.886 \pm 0.047$ & $415.4 \pm 0.2$\\
			$T_0\,\mathrm{[BJD - 2450000.0]}$ & $4812.47^{+0.30}_{-0.33}$ & $ 4812.52 \pm 0.36$ & $4813.42 \pm 0.24$\\
			$e$ & $0.8782 \pm 0.0025$ & $0.8847 \pm 0.0028$ & $0.856 \pm 0.009$\\
			$\omega\,[ ^\circ]$ & $1.2 \pm 1.0$ & $1.6 \pm 1.2$ & $352.9^{+1.9}_{-1.1}$\\
			$K\,[\ms]$ & $178.1^{+2.7}_{-2.6}$ & $181.1 \pm 3.7$ & $186.8 \pm 7.0$\\
			$a\,\mathrm{[AU]}$ (spec) & $1.091^{+0.068}_{-0.077}$ & $1.090 \pm 0.103$ & $1.149 \pm 0.017$\\
			$m_{\mathrm{P}}\,\sin i\,\mathrm{[M_J]}$ (spec) & $3.13^{+0.41}_{-0.43}$ & $3.11 \pm 0.42$ & $3.93^{+0.14}_{-0.15}$\\
			$a\,\mathrm{[AU]}$ (seis) & $1.187^{+0.050}_{-0.054}$ & $1.187 \pm 0.074$ & \textemdash\\
			$m_{\mathrm{P}}\,\sin i\,\mathrm{[M_J]}$ (seis) & $3.71^{+0.32}_{-0.33}$ & $3.68 \pm 0.33$ & \textemdash\\
			RMS about fit $[\ms]$ & $14.11$ & $13.79$ & $9.74$\\
			weighted RMS about fit $[\ms]$ & $11.71$ & $11.26$ & \textemdash\\\hline
			zero point (AAT) $[\ms]$ & $3.1^{+2.7}_{-2.5}$ & $2.5 \pm 4.4$ & $-$\\
			zero point (CHIRON) $[\ms]$ & $-68.4^{+1.9}_{-2.0}$ & $-74.2 \pm 5.5$ & $-$\\
			zero point (FEROS) $[\ms]$ & $-13.0 \pm 3.0$ & $-18.8 \pm 5.4$ & $-$\\
            zero point (FIDEOS) $[\ms]$ & $-93.9^{+4.3}_{-4.0}$ & $-94.1 \pm 7.3$ & $-$\\
			zero point (MJ1) $[\ms]$ & $-33.6^{+6.0}_{-4.6}$ & $-37.5 \pm 7.0$ & $-$\\
			zero point (MJ3) $[\ms]$ & $-54.3^{+4.9}_{-5.2}$ & $-61.6 \pm 8.1$ & $-$\\
			\hline
			$\sigma_{inst.jitt.}^2$ (AAT) $[m^2 s^{-2}]$ & $324.8^{+128.7}_{-89.2}$ & \textemdash & \textemdash\\
			$\sigma_{inst.jitt.}^2$ (CHIRON) $[m^2 s^{-2}]$ & $56.9^{+36.5}_{-26.9}$ & \textemdash & \textemdash\\
			$\sigma_{inst.jitt.}^2$ (FEROS) $[m^2 s^{-2}]$ & $14.6^{+43.8}_{-25.8}$ & \textemdash & \textemdash\\
			$\sigma_{inst.jitt.}^2$ (FIDEOS) $[m^2 s^{-2}]$ & $473.0^{+186.7}_{-132.3}$ & \textemdash & \textemdash\\
			$\sigma_{inst.jitt.}^2$ (MJ1) $[m^2 s^{-2}]^\dagger$ & $-225.7^{+34.7}_{-22.7}$ & \textemdash & \textemdash\\
			$\sigma_{inst.jitt.}^2$ (MJ3) $[m^2 s^{-2}]$ & $153.2^{+110.7}_{-81.7}$ & \textemdash & \textemdash\\
			\hline
		\end{tabular}
\end{table*}

\section{COMPANION UPPER MASS LIMIT}
\label{sec:mass_limit}
\subsection{Astrometric limit}
\label{sec:astrometric_limit}
\begin{figure}
	\includegraphics[width=\columnwidth,height=6.5cm]{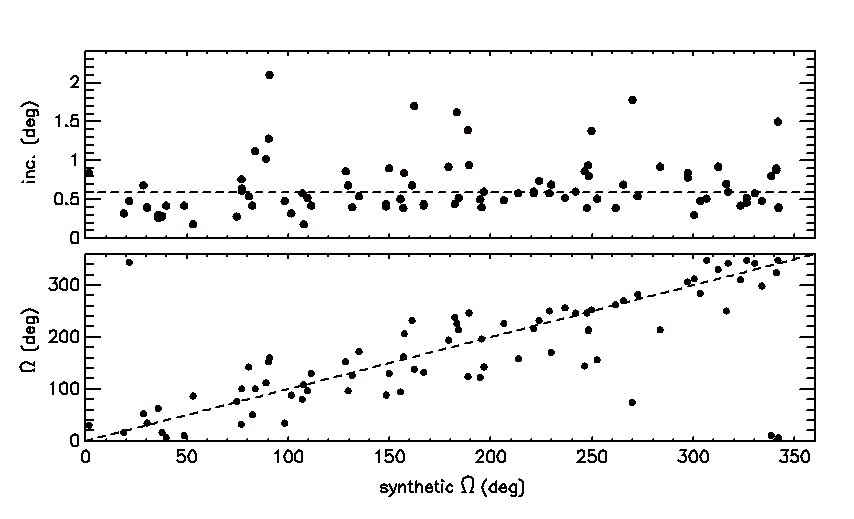}
	\caption{\textit{Upper panel}: Inclination angles recovered from the 100 simulations, as a function of synthetic $\Omega$ values. The horizontal dashed line corresponds to the input value of $i$ = 0.6 deg. \textit{Lower panel}: Same as the upper panel, but this time for the $\Omega$ values. The dashed line corresponds to the one-to-one correlation.}
	\label{fig:HIAD_sim}
\end{figure}

To better constrain the mass of HD\,76920\,b we applied the method 
presented in \citet{jones2017} (which is based on \citealt{sahlmann2011}), to derive the orbital inclination angle, and thus its dynamical mass. To do this, we combined the orbital elements derived here with the Hipparcos intermediate astrometric data (HIAD) obtained in \citet{new_hipred}. This dataset comprises a total of 106 one-dimensional abscissa measurements (see section 2.2.2 in \citet{new_hipred_2}), with a mean uncertainty of $1.9\,\mathrm{mas}$.
For this purpose, we first attempted to directly obtain a full 7-parameter orbital solution, solving for the inclination angle $i$ and the longitude of the ascending node $\Omega$, while keeping fixed the five parameters derived from the Keplerian fit ($P$, $e$, $\omega$, $K$, $T_0$), and correcting for the 5 single-star parameters solution ($\alpha^{\star}$, $\delta$, $\mu_{\alpha^{\star}}$, $\mu_{\delta}$, $\varpi$). Unfortunately, due to the small astrometric signal, in part due to the relatively small parallax (correspondingly large distance) of HD\,76920 ($\varpi$ = 5.41 $\pm$ 0.03 mas; \citet{gaia_dr2}), and also because of the high eccentricity (a significant astrometric perturbation occurs only close to periastron passage), no significant solution was obtained. This basically means that the orbital solution does not improve the standard Hipparcos 5-parameter solution. 
However, we could still compute an upper mass limit for the companion by injecting synthetic astrometric signals induced by the companion at different inclination angles (the smaller the value of $i$, the larger the astrometric signal). 
Briefly, we generated synthetic datasets, keeping fixed the time of the individual epochs of the HIAD and computing the expected astrometric signal induced by the companion by propagating the orbital solution to the epoch of the HIAD observations. This was performed at decreasing inclination angles, while randomly selecting $\Omega$ in the range between $0$--$360^{\circ}$. For each synthetic dataset we added Gaussian distributed uncertainties with standard deviation equal to the median abscissa error ($1.9\,\mathrm{mas}$ in this case).
Then we solved the 7-parameter solution to the synthetic datasets until we recovered the simulated ($i$, $\Omega$) pairs. Figure \ref{fig:HIAD_sim} shows the resulting ($i, \Omega$) values for a total of 100 synthetic datasets, with an input inclination angle of $0.6^{\circ}$, which corresponds to an angular semi-major axis of $1.9\,\mathrm{mas}$. As can be seen, for such a low $i$-value we are capable of recovering most of the synthetic signals with relatively good accuracy. We obtained a median of $i = 0.54^{\circ}$ with a standard deviation of $0.37^{\circ}$. We also note that only in eight cases we obtained a reduced $\chi^2$-value larger than for the Hipparcos 5-parameter solution. For comparison, we repeated this analysis for an inclination angle of $0.8^{\circ}$ (corresponding to $a = 1.5\,\mathrm{mas}$). We obtained a standard deviation of $0.73^{\circ}$, and already in 17\% of the simulations the reduced $\chi^2$ of the synthetic solution is larger than for the Hipparcos 5-parameter solution, showing how rapidly the detectability drops with the astrometric amplitude. In fact, only at inclination values of $i < 0.3^{\circ}$ we obtained reduced $\chi^2$-values of the synthetic solution lower than for the Hipparcos 5-parameter model in $> 99$\% of the cases. Based on this analysis, we might adopt a lower $i$-value of $\sim 0.4$--$0.6^{\circ}$, which corresponds to an upper mass limit for HD\,76920\,b of $\sim 0.4$--$0.6\,\mathrm{M_\odot}$. 
Finally, we note that by considering an orbital inclination angle of $90^{\circ}$ (edge-on orbit), the astrometric semi-major axis is only $20\,\mu\mathrm{as}$, which is comparable to the Gaia precision \citep{gaia}. It would thus be very challenging to significantly detect such a small signal even in the Gaia data.

\subsection{Geometric limit}
\label{sec:geometric}
The method described in Section\,\ref{sec:astrometric_limit} does not place very stringent upper limits on the mass of the planet. It is therefore interesting to note that we can put a much lower limit on the planet's mass from simple geometry. While the inclination is unknown, we know that there is no preferred orientation of the orbital plane with respect to the line of sight, in other words the inclination has an isotropic probability density function (pdf). This corresponds to a pdf that is flat in $\cos\,i$, which makes it easy to draw from for a Monte Carlo simulation using random inclination angles. We used a sample size of $10^8$ and found a 3-$\sigma$ (99.73\% confidence) upper mass limit of $42.6\,\mathrm{M_J}$, corresponding to an inclination of $4.2^{\circ}$. Results for a number of confidence levels are summarized in Table\,\ref{tab:geometric_limits}.

\begin{table*}
	\centering
	\caption{Geometric upper mass limits and corresponding inclinations for HD\,76920\,b for different confidence levels.}
	\label{tab:geometric_limits}
		\begin{tabular}{lcc}
		\hline confidence level & upper mass limit $[\mathrm{M_J}]$ & inclination $[^{\circ}]$ \\ \hline
		99.73\% & 42.6 & 4.2 \\
		99\% & 22.2 & 8.1 \\
		95\% & 10.0 & 18.2 \\
		90\% & 7.2 & 25.8 \\ \hline
		\end{tabular}
\end{table*}

\section{SUMMARY AND DISCUSSION}
\label{sec:discussion}
We obtained 63 new radial velocity measurements of HD\,76920 from four different instruments around the time of the predicted periastron passage. The unusually high eccentricity of HD\,76920\,b means that $\sim 90$\% of the peak-to-peak RV is traversed up and down in only $\sim 14$\% of the orbital period, and the RV curve is approximately flat for the remaining $\sim 360$ days of the orbital period, making it difficult to determine precise orbital elements from an observational point of view. However, in order to constrain the orbital elements, it is essential to have good sampling of the non-flat parts of the orbit where the RV changes rapidly over time. As the orbital phase near periastron passage was not very well covered in their initial work, \citet{w17} suggested follow-up observations be carried out during the next periastron passage. Flexible scheduling of observing time during periastron passage on telescopes with high-resolution spectrographs is an effective way of confirming the nature of highly eccentric planets and determining their orbital properties and minimum mass to high precision. For example, HD\,37605\,b with an eccentricity of $e = 0.74$, or HD\,45350\,b with an eccentricity of $e = 0.76$ have been confirmed in this way \citep{cochran2004, endl2006}.

\begin{figure}
	\includegraphics[width=\columnwidth]{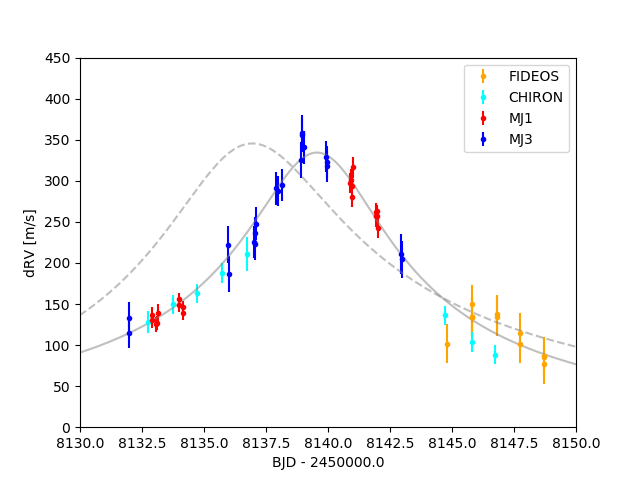}
	\caption{Comparison of the best-fit model RV curve from this work (solid grey line) with the one from \citet{w17} (dashed grey line). The periastron passage happened about 3 days later than predicted by the orbital elements from \citet{w17}.}
	\label{fig:peaktimes}
\end{figure}

We were fortunate enough to be granted access to the HERCULES, CHIRON, FEROS, and FIDEOS spectrographs during that period, and hence managed to obtain near-continuous coverage of the corresponding RV peak. In hindsight, getting enough telescope time either side of the predicted periastron passage was particularly important because the periastron passage actually happened about 3 days later than predicted, or about $3.7$-sigma when we calculate the uncertainty on the time of periastron passage via bootstrapping based on the orbital elements given by \citet{w17} (see Fig.~\ref{fig:peaktimes}). This highlights the importance and scientific value of small-to-medium size telescopes with high-resolution spectrographs to the exoplanet community \citep[e.g.][]{swift15, minervasouth}, as it would have been near impossible to get enough time on larger telescopes at such short notice.

Our best-fit orbital solution to the combined data set from a total of five instruments (and six different operating modes) yielded an even higher than expected eccentricity of $e=0.88$ and an orbital period of $415.9\,\mathrm{d}$. We also found a RV semi-amplitude of $178\,\ms$, and a semi-major axis of $1.09\,\mathrm{AU}$. The new orbital solution corresponds to minimum mass of $3.1\,\mathrm{M_J}$ for the planet that is about 20\% lower than that reported by \citet{w17}, mainly owing to our new lower stellar mass estimate. Formally, the RMS of the residuals from our fit is larger than in \citet{w17}, partly because the individual uncertainties in and the scatter of the HERCULES data are about three times larger compared to the other instruments, and partly because we have effectively given different weights to the RV measurements during the fitting via the error treatment described in section \ref{sec:solution}. In particular, note that the RMS of the AAT residuals from our best fit is now $17.0\,\ms$, which is of course expected as they now carry less weight compared to \citep{w17}. More importantly though, due to the much improved phase coverage near periastron passage the uncertainties in the orbital elements are now significantly reduced. Notably, the uncertainty in the orbital period was reduced by a factor of 5, and the uncertainty in the eccentricity was reduced by more than a factor of 3. In addition, we also estimated an upper mass limit of $0.4$--$0.6\,\mathrm{M_\odot}$ for the companion from HIPPARCOS astrometry, and a 3-$\sigma$ upper mass limit of $42.6\,\mathrm{M_J}$ from geometric considerations.

\subsection{Star--Planet Interactions}
Note that our value of the semi-major axis is slightly smaller the one given by \citet{w17}, while our eccentricity is slightly larger. However, this seemingly small difference means that the planet actually comes to within $2.4 \pm 0.3$ stellar radii from its host star's surface at closest approach, or about one stellar radius closer than estimated by \citet{w17}. Also note that we calculate the same value for planet's distance from the stellar surface at periastron in units of the stellar radius, independent of whether we use the spectroscopic or the asteroseismic stellar parameters. Naturally, this makes the system a prime target for studies of star--planet interactions.

The orbital evolution is determined by the combined effect of tidally induced orbital decay and mass-loss induced orbital expansion. Following the same procedure as in \citet{V14} and \citet{w17}, we determined the planet's orbit and eccentricity evolution as the star evolves up the Red Giant Branch, and the tidal dissipation is dominated by motions in the star's convective envelope \citep{zahn77}. We have updated both the planetary and stellar parameters since \citet{w17}. We use \texttt{SSE} stellar models \citep{hurley2000} with the asteroseismic mass of $1.3\,$\msun\, and a Solar metallicity. We furthermore consider two values for the Reimers $\eta$ mass loss parameter \citep{reimers_eta}: a standard value of $\eta=0.6$, and an extreme case of $\eta=0.0$ (no mass loss). The latter means that the planet experiences only tidal decay on its orbit, not any orbital expansion owing to mass loss, and hence represents an optimal case for a maximum orbital decay owing to tidal forces.

In each case, the planet's orbit undergoes only very minor decay before the planet is engulfed in the stellar envelope: for both $\eta=0.6$ and $\eta=0.0$, the orbital eccentricity decays by only about $0.002$ before the star engulfs the planet. With no stellar mass loss ($\eta=0.0$), the semi-major axis decays by $0.01$\,AU, while with mass loss ($\eta=0.6$) the semi-major axis increases by $0.003$\,AU. The planet enters the stellar envelope when the star has grown to a little over 3 times its present radius, some 50\,Myr hence. Adopting the spectroscopic stellar mass of $1.0\,$\msun does not qualitatively change the future evolution: the eccentricity decay is a little larger ($0.004-0.005$) and the remaining lifetime a little longer ($\sim 80$\,Myr), but there are still no large changes to the orbit expected.

\subsection{Transit Probability}
\begin{figure}[ht]
	\includegraphics[width=\columnwidth]{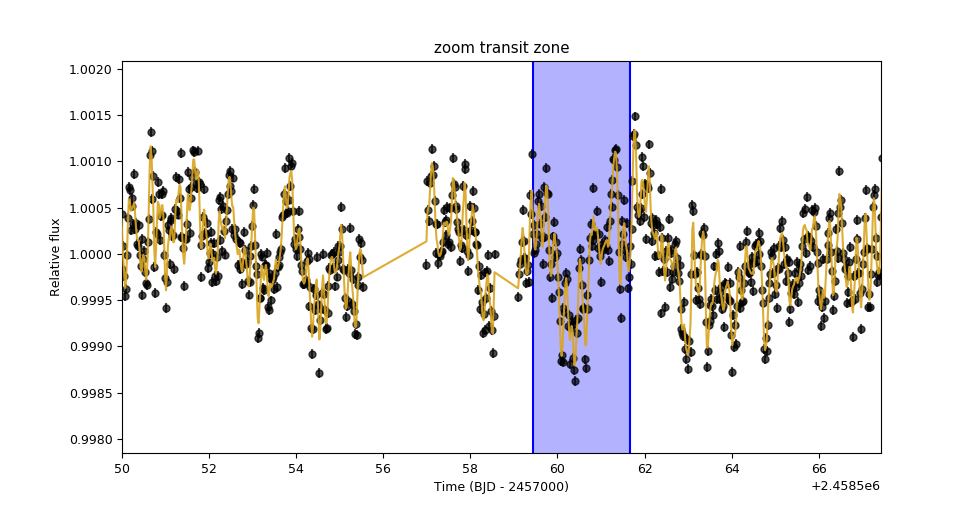}	\caption{TESS light curve around the expected transit time, which is highlighted in light blue. The yellow line represents the Gaussian process fit.}
	\label{fig:curvezoom}
\end{figure}

\begin{figure}[ht]
	\includegraphics[width=\columnwidth ]{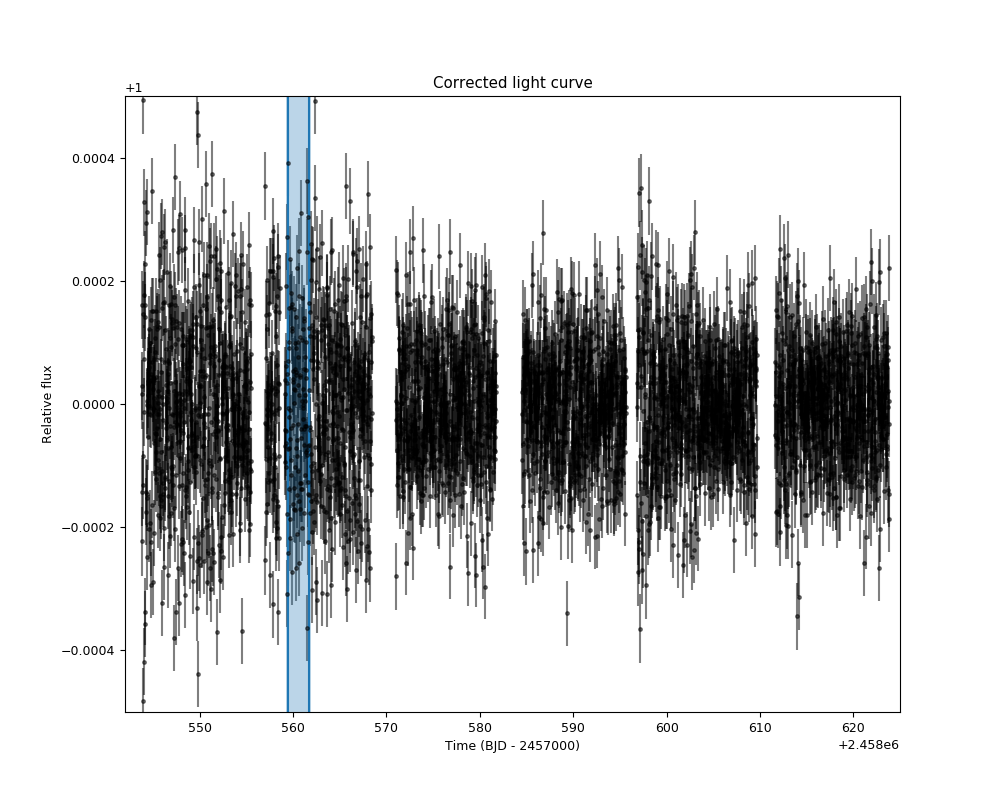}
	\caption{GP corrected TESS light curve of HD\,76920. The expected transit time is highlighted in light blue.}
	\label{fig:curveGP}
\end{figure}

Our updated orbital solution also leads to an even higher transit probability than the $10.3\%$ reported in \citet{w17}. From the \texttt{emcee} best-fit orbital elements listed in Table \ref{tab:orbel}, it follows that at inferior conjunction, which happens at a true anomaly of $88.8^{\circ}$ and $5.06\,\mathrm{d}$ after periastron passage, the star-planet separation is $0.245\,\mathrm{AU}$, and the azimuthal component of the orbital velocity is $60.7\,\mathrm{km\,s^{-1}}$. In order to calculate the probability, depth, and duration of a potential transit, we must first have an estimate of the planetary radius. We used the mass-radius relationship for the Jovian regime in form of the power law $R_{pl} \propto m_P^{-0.04}$ as given by \citep{chenkipping2017}, with which we calculated the planet's radius to be $0.96\,R_{J}$. With that in hand, we calculated a relatively high transit probability of $16.0\%$, using Eq.~5 from \citet{kane2008}, with a duration of $2.2\,\mathrm{d}$ and a transit depth of $0.013\,\%$ or 130 ppm, assuming an inclination of $i=90^{\circ}$ and ignoring limb darkening effects. While the large stellar radius increases the transit probability, unfortunately it also decreases the transit depth, requiring a level of photometric precision that is extremely challenging for ground-based transit searches, albeit perhaps not impossible \citep[e.g.][]{wasp_precision_2013}. However, TESS can technically achieve the required precision for a star of this magnitude \citep{tess}. HD\,76920 has ecliptic coordinates of about $\lambda = 202.5^{\circ}$ and $\beta = -73.2^{\circ}$, and therefore lies about five degrees outside the southern TESS continuous viewing zone. However, by pure coincidence, this placed HD\,76920 inside the sector that TESS was observing at the time of the next potential transit (sector 9), which we predicted to occur approximately between JD2458559.43$\,\pm\,$0.63 and JD2458561.66$\,\pm\,$0.63 (UT $16.93 - 19.16$ March 2019). We list predicted mid-transit times, as well as ingress and egress times for potential future transits in Table \ref{tab:transits}. \newline \indent
Unfortunately, due to it being an evolved star, HD\,76920 presents a photometric variability at the 500 ppm level, which is significantly larger than the expected transit depth. However, before searching for a potential transit signal, we corrected the light curve using a Gaussian process (GP), following a similar procedure to that described in \citet{jones2019}. The fit was performed with the \texttt{Juliet} code \citep{juliet2019} using a Mat\'ern kernel. To model the asteroseismic
 signal we used a Gaussian prior with mean equal to the period corresponding to $\nu_{max}$, as derived in section \ref{sec:asteroseismology}. Figure \ref{fig:curvezoom} shows the TESS light curve and the GP model. \newline \indent
 Finally, to determine if the planet transits on the predicted date, we compared the Bayesian evidence between a transit model and a non-transit model. As we found no significant difference, we assumed the simpler model, i.e. the non-transit model. The GP corrected light curve around the expected transit time is shown in Figure \ref{fig:curveGP}.

\begin{table*}
	\centering
	\caption{Predicted windows for potential past and future transits of HD\,76920\,b.}
	\label{tab:transits}
		\begin{tabular}{cccc}
		\hline ingress & mid-transit & egress & UT date (mid)\\
		& [JD - 2400000.0] &\\ \hline
		$58559.43 \pm 0.63$ & $58560.55 \pm 0.54$ & $58561.66 \pm 0.63$ & 18 Mar 2019\\
		$58975.32 \pm 0.66$ & $58976.44 \pm 0.58$ & $58977.56 \pm 0.66$ & 06 May 2020\\
		$59391.21 \pm 0.68$ & $59392.33 \pm 0.61$ & $59393.45 \pm 0.68$ & 26 Jun 2021\\
		$59807.10 \pm 0.71$ & $59808.22 \pm 0.64$ & $59809.34 \pm 0.71$ & 16 Aug 2022\\
		$60222.99 \pm 0.75$ & $60224.11 \pm 0.68$ & $60225.23 \pm 0.75$ & 06 Oct 2023\\
		\hline
		\end{tabular}
\end{table*} 

\subsection{Origin of the high eccentricity}
The origin of Hot Jupiters and/or highly eccentric planets is usually explained via the Kozai-Lidov mechanism, whereby perturbations caused by a massive third body (i.e. a stellar companion) can cause oscillations between the planet's eccentricity and inclination as long as its angular momentum component parallel to the orbital angular momentum of the two stars remains constant \citep{lidov1962, kozai1962}. However, as already noted by \citet{w17}, there are no indications for additional massive companions in the RV data. Also, while the Gaia DR2 catalogue \citep{gaia, gaia_dr2} lists two faint $(G \sim 21\,\mathrm{mag})$ stellar objects (Gaia DR2 IDs 5224124753994137984 and 5224127812011479808) with compatible parallaxes within 5 arcmins from HD\,76920, corresponding to a physical separation of $\lesssim 55\,000\,\mathrm{AU}$ at a distance of $d=184\,\mathrm{pc}$ \citep{bailerjones2018}, their respective proper motions and $B-R$ colours are very different from the corresponding values for HD\,76920, which rules them out as physically close companions. Furthermore, as also noted by \citet{w17}, in the Kozai-Lidov scenario planets like HD\,76920\,b are readily being engulfed by their host stars as they move up the red giant branch \citep{fh2016}, and, according to simulations by \citet{parker2017}, free-floating planets are almost exclusively captured into much wider orbits $(a > 100\,\mathrm{AU})$. This leaves planet-planet scattering as the most likely explanation for the highly eccentric orbit of HD\,76920\,b \citep[e.g.][]{pp_scattering}. In this scenario, a second planet of comparable mass would have either been ejected from the system as the result of a close encounter with HD\,76920\,b, or at least pushed outwards into a long-period orbit that is beyond our current detection limits. A third option is that the second planet disappeared from the system because it was thrown towards the star and engulfed by it.

\subsection{Additional Considerations}
The very high eccentricity of HD\,76920\,b not only makes it the most eccentric planet known to orbit an evolved star by some margin, but also puts it in fifth place amongst all known exoplanets\footnote{\texttt{http://exoplanet.eu} \citep{exoplanet_eu}. Note that when last accessed on 31 January 2021, WASP-74\,b was erroneously listed to have an eccentricity of 0.88, but it really is the limb-darkening coefficient $\epsilon$ that has this value \citep{wasp74b_rm}.}. Its eccentricity is only surpassed by HD\,4113\,A\,b ($e=0.90$) \citep{hd4113b}, HD\,7449\,A\,b ($e=0.92$) \citep{dumusque2011, truly_eccentric1}, HD\,80606\,b ($e=0.93$) \citep{hd80606b, hd80606b_update}, and HD\,20782\,b ($e=0.97$) \citep{hd20782b, hd20782b_update, kane16}, all of which are gas giants orbiting solar-mass main-sequence stars. 

Several studies have highlighted the possibility that a RV curve produced by two low-eccentricity planets can be misinterpreted as being caused by one planet with medium to high eccentricity, especially for low signal-to-noise ratios $K/ \sigma$, poor sampling, and/or if the two planets are in resonant orbits \citep[e.g.][]{shenturner2008, rh2009, anglesc2010, rob2012, rob2013, truly_eccentric1, truly_eccentric2}. However, given the large $K/ \sigma$ ratio, the dense sampling we have achieved around periastron passage, and the very high eccentricity (which is well outside the "danger zone" as identified by \citet{truly_eccentric2}, i.e. the range of eccentricities that can be most easily mimicked by two near-circular planets), we are very confident that the results presented in this paper remove any possibly remaining doubts about the RV variations being caused by a single planetary companion in a highly eccentric orbit.

\begin{acknowledgements}
CB was supported by Australian Research Council Discovery Grant DP170103491. LV acknowledges support from CONICYT through projects Fondecyt n. 1171364 and Anillo ACT-1417. AZ is supported by CONICYT grant n. 2117053. SW thanks the Heising-Simons Foundation for their generous support. AJM acknowledges support from the Knut and Alice Wallenberg Foundation (project grant 2014.0017), the Swedish Research Council (starting grant 2017-04945), and the Walter Gyllenberg Foundation of the Royal Physiographic Society of Lund.
RB acknowledges support from FONDECYT Project 11200751, from CORFO project N$^\circ$14ENI2-26865, and from project IC120009 ``Millennium Institute of Astrophysics (MAS)'' of the Millenium Science Initiative, Chilean Ministry of Economy.
We are grateful for receiving a generous allocation of observing time at UCMJO. This research has made use of NASA's Astrophysics Data System (ADS), and the SIMBAD database, operated at CDS, Strasbourg, France. This research has also made use of the Extrasolar Planets Encyclopaedia at \texttt{http://www.exoplanet.eu}. This work has made use of data from the European Space Agency (ESA) mission {\it Gaia} (\url{https://www.cosmos.esa.int/gaia}), processed by the {\it Gaia} Data Processing and Analysis Consortium (DPAC, \url{https://www.cosmos.esa.int/web/gaia/dpac/consortium}). Funding for the DPAC has been provided by national institutions, in particular the institutions participating in the {\it Gaia} Multilateral Agreement. Finally, we would like to thank the anonymous referee for their insightful comments that helped noticeably to improve this manuscript.\\
\end{acknowledgements}

\begin{appendix}

\section{RADIAL VELOCITY DATA}

\begin{table}
    \caption{Radial velocities for HD\,76920. Note that the velocities shown have instrument-specific zero points, which are included in the fitting process and are given in Table \ref{tab:orbel}. The AAT data was taken from \citet{w17}.}
	\centering
	\label{tab:rvs}
	\begin{tabular}{lrrr}
		\hline \hline
		BJD-2450000.0 & RV $[\ms]$ & $\sigma_{\mathrm{RV,int}}$ $[\ms]$ &
		Instr.\\
		\hline
		4867.07428  &  17.9 &  2.2 & AAT \\
		5226.21880  & 269.5	&  5.3 & AAT \\
		5227.20104  & 303.4	&  3.7 & AAT \\
		5318.89227  &   1.0	&  1.9 & AAT \\
		5602.04422  &   6.1	&  1.9 & AAT \\
		5880.22005  & -55.3	&  2.3 & AAT \\
		5906.11204  & -31.3	&  1.8 & AAT \\
		5907.19640  & -28.1	&  2.6 & AAT \\
		5969.07596  & -15.5	&  2.1 & AAT \\
		6088.86366  &  54.1	&  3.8 & AAT \\
		6344.02991  &  -3.1	&  2.7 & AAT \\
		6374.98803  & -16.4	&  2.4 & AAT \\
		6376.95955  & -14.1	&  2.4 & AAT \\
		6377.96197  & -25.2	&  2.6 & AAT \\
		6399.96882  & -18.5	&  3.1 & AAT \\
		6530.31941  &  11.0	&  3.0 & AAT \\
		6744.98572  &  -7.3	&  2.4 & AAT \\\hline
		7306.82769   & 264.5	&  4.4 & CHIRON \\
		7324.78909   & -16.0	&  4.5 & CHIRON	\\
		7365.78945   & -66.0	&  4.1 & CHIRON	\\
		7433.69902   & -97.2	&  3.6 & CHIRON	\\
		7433.71312   & -99.6	&  3.4 & CHIRON	\\
		7433.72722   & -94.7	&  3.4 & CHIRON	\\
		7458.68833   & -90.4	&  3.3 & CHIRON	\\
		7458.70243   & -95.2	&  3.7 & CHIRON	\\
		7458.71651   & -98.4	&  3.6 & CHIRON	\\
		7478.64626   & -76.3	&  4.1 & CHIRON	\\
		7478.66036   & -79.6	&  3.7 & CHIRON	\\
		7478.67445   & -77.2	&  3.7 & CHIRON	\\
		8132.74057   &  60.5	&  8.0 & CHIRON	\\
		8133.75634   &  81.7	&  6.2 & CHIRON	\\
		8134.73750   &  95.1	&  5.0 & CHIRON	\\
		8135.72967   & 120.3	&  7.0 & CHIRON	\\
		8136.74525   & 143.4	& 18.1 & CHIRON	\\
		8144.69957   &  68.5	&  6.0 & CHIRON	\\
		8145.80543   &  36.0	&  6.9 & CHIRON	\\
		8146.73179   &  20.7	&  5.3 & CHIRON	\\\hline
		7641.91298  & -40.6 &  4.0 & FEROS \\
		7643.90565  & -25.6 &  5.1 & FEROS \\
		7700.84513  &  20.7 &  5.4 & FEROS \\
		7702.87010  &  19.1 &  4.3 & FEROS \\
		7703.79908  &  33.4 &  4.5 & FEROS \\
		7705.85500  &  36.7 &  4.3 & FEROS \\
		7894.56042  & -38.1 &  4.2 & FEROS \\
		7895.46977  & -35.7 &  3.8 & FEROS \\
		8123.84502  &  30.5 &  4.1 & FEROS \\\hline
	\end{tabular}
\end{table}

\begin{table}
\label{tab:continued}
\begin{tabular}{lrrr}
		\hline \hline
		BJD - 2450000.0 & RV $[\ms]$ & $\sigma_{\mathrm{RV,int}}$ $[\ms]$ & Instrument\\
		\hline
        8101.764230 & -91.1 & 6.4 & FIDEOS \\
		8111.810817 & -75.7 & 5.9 & FIDEOS \\ 
		8113.812973 & -74.9 & 7.6 & FIDEOS \\
		8144.796552 &   6.0 & 7.2 & FIDEOS \\
 		8145.804861 &  38.7 & 6.9 & FIDEOS \\
		8145.811483 &  38.0 & 6.8 & FIDEOS \\
		8145.818042 &  38.2 & 7.3 & FIDEOS \\
		8145.825082 &  53.7 & 7.1 & FIDEOS \\
		8146.795174 &  38.5 & 6.3 & FIDEOS \\
        8146.802702 &  42.0 & 6.2 & FIDEOS \\
		8147.748904 &   5.8 & 6.4 & FIDEOS \\
		8147.756478 &  19.0 & 6.6 & FIDEOS \\
		8148.710383 &  -9.9 & 6.6 & FIDEOS \\
		8148.718152 &  -9.4 & 6.2 & FIDEOS \\
		8148.725379 & -19.2 & 7.3 & FIDEOS \\\hline
		8132.894243	&  101.3 & 16.4	& MJ1 \\
		8132.916154	&   94.8 & 16.6	& MJ1 \\
		8133.082251	&   89.7 & 16.1	& MJ1 \\
		8133.104068	&   91.0 & 15.6	& MJ1 \\
		8133.129350	&  103.0 & 17.4	& MJ1 \\
		8133.985792	&  112.8 & 15.5	& MJ1 \\
		8134.008329	&  119.8 & 15.5	& MJ1 \\
		8134.142827	&  103.6 & 15.7	& MJ1 \\
		8134.165035	&  110.1 & 15.7	& MJ1 \\
		8140.904009	&  261.4 & 18.3	& MJ1 \\
		8140.926304	&  265.6 & 18.4	& MJ1 \\
		8140.948490	&  245.1 & 18.3	& MJ1 \\
		8140.970459	&  257.5 & 19.7	& MJ1 \\
		8140.993368	&  280.5 & 18.4	& MJ1 \\
		8141.920501	&  226.1 & 17.5	& MJ1 \\
		8141.940818	&  220.8 & 18.9	& MJ1 \\
		8141.968963	&  221.5 & 18.3	& MJ1 \\
		8141.990759	&  226.9 & 15.7	& MJ1 \\
		8142.013180	&  206.3 & 17.4	& MJ1 \\\hline
		8131.973084	&   77.4 & 14.4	& MJ3 \\
		8131.995427	&   59.8 & 12.4	& MJ3 \\
		8135.979715	&  166.9 & 17.1	& MJ3 \\
		8136.002019	&  131.8 & 16.6	& MJ3 \\
		8137.019531	&  170.2 & 13.7	& MJ3 \\
		8137.041852	&  181.2 & 13.3	& MJ3 \\
		8137.064108	&  168.3 & 13.8	& MJ3 \\
		8137.089041	&  192.7 & 14.5	& MJ3 \\
		8137.919896	&  235.7 & 14.3	& MJ3 \\
		8137.974526	&  232.1 & 11.8	& MJ3 \\
		8138.133998	&  240.0 & 13.5	& MJ3 \\
		8138.922459	&  269.9 & 16.4	& MJ3 \\
		8138.944974	&  303.4 & 16.7	& MJ3 \\
		8138.969931	&  300.3 & 14.2	& MJ3 \\
		8139.032653	&  285.9 & 14.2	& MJ3 \\
		8139.912303	&  274.3 & 12.9	& MJ3 \\
		8139.944173	&  268.1 & 13.2	& MJ3 \\
		8139.966157	&  262.9 & 13.5	& MJ3 \\
		8142.931486	&  155.6 & 20.3	& MJ3 \\
		8142.975638	&  149.0 & 17.9	& MJ3 \\
		\hline \hline
	\end{tabular}
\end{table}

\section{Corner plot of the fit parameters}
\begin{figure*}
	\includegraphics[width=2.0\columnwidth]{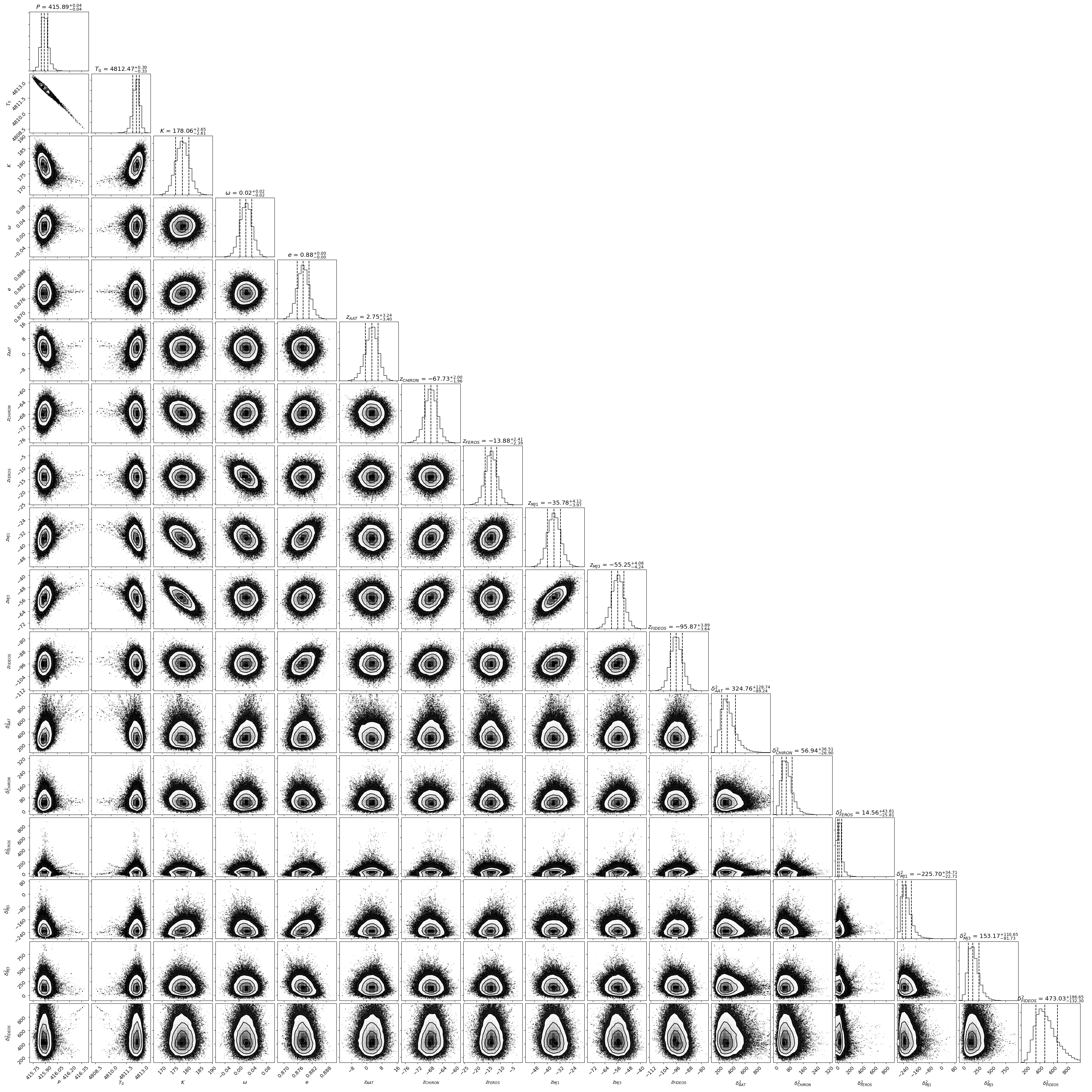}
	\caption{Corner plot of the posterior probability distributions of the 17 free parameters used in the \texttt{emcee} fitting. All numerical values shown are rounded to two decimal places.}
	\label{fig:corner_plot}
\end{figure*}
 
\end{appendix}

\bibliographystyle{pasa-mnras}
\bibliography{quellen}

\end{document}